\documentclass[10pt,a4paper,english]{article}
\usepackage{color}
\usepackage[T1]{fontenc}
\usepackage[latin9]{inputenc}
\usepackage{amsfonts}
\usepackage{amssymb}
\usepackage{mathrsfs}
\usepackage{amsthm}
\usepackage{graphicx}
\usepackage{amsmath}
\newcommand{\BHH}{{\mathcal{B}_H}}
\newcommand{\im}{\mathrm{Im}}
\newcommand{\BH}{{\mathcal{B}(\mathcal{H})}}

\newcommand{\ra}{{\, \rightarrow\, }}
\newtheorem{thm}{Theorem}
\newtheorem{con}{Conjecture}
\newtheorem{prop}{Proposition}
\newtheorem{cor}{Corollary}

\theoremstyle{definition}

\newtheorem{remark}{Remark}

\newcommand{\Tr}[0]{\mathrm{Tr}}

\newcommand{\bei}{\begin{itemize}}
\newcommand{\eei}{\end{itemize}}

\def\<{\langle}
\def\>{\rangle}
\newcommand{{\Cn}}{{\mathbb{C}^n}}
\newcommand{{\CN}}{{\mathbb{C}^{2n}}}
\newcommand{{\BC}}{{\mathcal{B}(\mathbb{C}^n)}}
\newcommand{{\BBC}}{{\mathcal{B}(\mathbb{C}^{2n})}}

\begin{document}

  \title{\textbf{Generalising Wigner's theorem}}

  \author{Gniewomir Sarbicki and Dariusz  Chru\'sci\'nski\\
  Institute of Physics, Faculty of Physics, Astronomy and Informatics \\ Nicolaus Copernicus University,\\
  Grudzi\c{a}dzka 5/7, 87--100 Toru\'n, Poland \\ \\
  Marek Mozrzymas \\
Institute for Theoretical Physics, University of Wroc{\l}aw \\ 50-204 Wroc{\l}aw, Poland}

  \maketitle

  \begin{abstract}
    We analyse linear maps of operator algebras $\BHH(\mathcal{H})$ mapping the set of rank-$k$ projectors onto the set of rank-$l$ projectors surjectively. A complete characterisation of such maps for prime $n = \dim\mathcal{H}$ is provided. A particular case corresponding to  $k=l=1$ is well known as the Wigner's theorem. Hence our result may be considered as a generalization of this celebrated Wigner's result.
  \end{abstract}

\section{Introduction}

The celebrated Wigner's theorem \cite{Wigner} in its original formulation  says that any map $\Phi$ between rank-1 projectors in a Hilbert space preserving the Hilbert-Schmidt product, i.e. $(\Phi(P_1),\Phi(P_2))_{\rm HS} = (P_1,P_2)_{\rm HS}$, is of the form:
\begin{equation}\label{!}
  \Phi(X) = U X U^\dagger \ \ \mbox{or} \ \   \Phi(X) = U X^{\rm t} U^\dagger \ ,
\end{equation}
where $X^{\rm t}$ denotes a transposition with respect to a fixed orthonormal basis in $\mathcal{H}$ and $U$ is a unitary operator (see also \cite{Lahti} and recent analysis in \cite{Simon}). It is clear that any such map induces unitary or antiunitary operation in the original Hilbert space. 
The Wigner's theorem is sometimes reformulated as follows \cite{Werner}: one restricts to linear maps $\Phi : \BH \ra \BH$ which are not necessarily  Hilbert-Schmidt isometries. Now, any such map that  maps bijectively rank-1 projectors to rank-1 projectors is of the form (\ref{!}).
Clearly, a map mapping  rank-1 projectors to rank-1 projectors is by construction a positive map \cite{Paulsen,Bhatia} and hence the Wigner's theorem states that a  positive trace-preserving map $\Phi$ has a positive inverse if and only it has a form (\ref{!}).

In this paper we consider linear maps mapping surjectively rank-$k$ projectors to rank-$l$ projectors. We  show for which $k,l$ such maps exist and that for prime dimensions of the Hilbert space they has exactly the form (\ref{!}). We also provide an example of a map which is not of the form (\ref{!}). Interestingly, linear maps acting surjectively between sets of projectors of fixed rank maps have recently attracted attention in the problem of entanglement detection \cite{Adam}.



Let us consider a real space of self-adjoint operators $\BHH(\mathcal{H})$.  We denote the set of rank-$k$ projectors supported on a subspace $V \subset \mathcal{H}$ by $\mathcal{P}_k(V) \subset \BHH$. It is a smooth manifold of dimension $k(n-k)$. Let $\Phi$ be a linear endomorphism of $\BH$ and let $\Phi(\mathcal{P}_k(\mathcal{H}))=\mathcal{P}_l(\mathcal{H})$. From the fact, that $\mathcal{P}_l(\mathcal{H})$ spans the whole $\BH$, one gets that $\Phi$ is a surjective endomorphism, hence a bijection, hence also a bijection between sets $\mathcal{P}_k(\mathcal{H})$ and $\mathcal{P}_l(\mathcal{H})$. This implies, that $l=k$ or $l=n-k$.
There is a natural linear isomorphism between $\mathcal{P}_k(\mathcal{H})$ and $\mathcal{P}_{n-k}(\mathcal{H})$:
\begin{equation}\label{R}
  R_k(X) = \frac 1k \,\mathbb{I}\, \Tr X - X \ .
\end{equation}
Hence any bijection $\Phi:\mathcal{P}_k(\mathcal{H}) \to \mathcal{P}_{n-k}(\mathcal{H})$ can be represented as a composition of $R_k$ and bijective endomorphism of $\mathcal{P}_k(\mathcal{H})$. Therefore,  it suffices to characterise all bijective endomorphisms of $\mathcal{P}_k(\mathcal{H})$.

Note, that this problem belongs to so-called linear preserver problem which deal with
characterization of linear operators  that leave certain properties
or certain subsets in its domain
invariant. This program was started already by Frobenius \cite{Frobenius}. A well known example of linear preservers are rank preserver, nilpotency preserver and spectrum preserver: $\Phi$ is rank preserver iff  $\Phi(X) = MXN$ or $\Phi(X) = MX^{\rm t}N$, where $M,N$ are invertible elements from $M_n(\mathbb{C})$. $\Phi$ provides a  preserver of nilpotency iff $\Phi(X) = MXN$ or $\Phi(X) = MX^{\rm t}N$, where $M,N \in M_n(\mathbb{C})$ such that $MN = c \mathbb{I}_n$ and $c \in \mathbb{C}$. Finally, $\Phi$ is a spectrum preserver iff $\Phi(X) = M X M^{-1}$ or $\Phi(X) = M X^{\rm t} M^{-1}$.

Clearly, $\Phi$ defined in (\ref{!}) is an example of rank preserver and nilpotency preserver, but in our problem the restriction is weaker - we demand preserving the nilpotency only for a one given value of the rank (the set of nilpotent hermitian operators splits into connectivity components grouping the projectors of the same rank). With a weaker assumption one can expect that the resulting set of linear operation can be in general greater. Indeed, for a special choice of $n$ and $k$ we provide an example of an invertible map preserving rank-$k$ projector not being of the form (\ref{!}).

Recently Marciniak \cite{MM}  considered a related problem and shown that every positive map $\Phi$ such that ${\rm rank} \Phi(P) \leq 1$ for any rank-1 projector $P$ is the rank-1 preserver and  has a form (\ref{!}).

\section{Main result}

This section provides the main result of the paper. Let us start with the following 

\begin{prop} \label{ort}
   Any $\Phi: \BHH(\mathcal{H}) \to \BHH(\mathcal{H})$ mapping bijectively $\mathcal{P}_k(\mathcal{H})$ onto itself preserves the orthogonality.

   \proof If $2k>n$, then there are no non-zero mutually orthogonal elements in $\mathcal{P}_{k}(\mathcal{H})$ and the proposiion is true in a trivial way. We will consider the case when $2k \le n$.  Let $P_V$ be the projector onto a $2k$-dimensional subspace $V \subset \mathcal{H}$. $P_V$ can be decomposed in various ways into the sum of two rank-$k$ orthogonal projectors $P_1$ and $P_2$. They are mapped via $\Phi$ onto two rank-$k$ projectors $Q_1=\Phi(P_1)$ and $Q_2=\Phi(P_2)$. From the positivity of $Q_1$ and $Q_2$ one has $\im Q_i = \im \Phi(P_i) \subset \im \Phi(P_V)$. One can repeat it for any choose of $P_1$ and $P_2$, hence $\forall P \in \mathcal{P}_k(V) \ \im \Phi(P) \subset \im\Phi(P_V)$ and thus $\Phi(\mathcal{P}_k(V)) \subset \BHH(\mathrm{Im}\Phi(P_V))$. Because $\mathcal{P}_k(V)$ spans the whole $\BHH(V)$, we have that $\Phi(\BHH(V)) \subset \BHH(\im\Phi(P_V))$.

   While $\Phi$ is bijective on $\mathcal{P}_k(\mathcal{H})$, it is bijective on $\mathcal{B}_H(\mathcal{H})$. Hence
   $\dim \BHH(V) \le \dim \BHH(\im\Phi(P_V))$ and hence $\dim V \le \dim \im\Phi(P_V)$.
   While $\Phi(P_V)$ is a sum of two rank-$k$ projectors, $\dim \im\Phi(P_V) \le 2k = \dim V$ and hence $\dim \im\Phi(P_V) = \dim V$. $\Phi$ establishes a bijection between $\BHH(V)$ and $\BHH(\im\Phi(P_V))$ and hence between $\mathcal{P}_k(V)$ and $\mathcal{P}_k(\im\Phi(P_V))$. Any rank-$k$ projector $Q \in \mathcal{P}_k(\im\Phi(P_V))$ can be realised as $\Phi(P)$ for some $P \in \mathcal{P}_k(V)$.

   Now we choose the basis of $\im \Phi(P_V)$ to make $\Phi(P_V)$ diagonal. Take a rank-$k$ dimensional projector $Q \in \mathcal{P}_k(\im\Phi(P_V))$, diagonal in this base (commuting with $\Phi(P_V)$). The operator $\Phi(P_V)-Q = \Phi(P_V)-\Phi(P) = \Phi(P_V-P)$ is a rank $k$-projector, diagonal in the chosen basis. One can easily find, that this implies, that only possible values on the diagonal of $\Phi(P_V)$ are $1$s and $2$s. But the rank of $\Phi(P_V)$ and its trace are equal $2k$, so $\Phi(P_V)$ is the projector onto $\im \Phi(P_V)$.

   For any two orthogonal projectors $P_1$ and $P_2$ the operator $\Phi(P_1 + P_2) = \Phi(P_1) + \Phi(P_2)$ is a rank-$2k$ dimensional projector, hence the projectors $\Phi(P_1)$ and $\Phi(P_2)$ are orthogonal \hfill $\square$

  \end{prop}

  One has immediately the following:

  \begin{cor} \label{qk}
   Any $\Phi: \BHH(\mathcal{H}) \to \BHH(\mathcal{H})$ mapping bijectively $\mathcal{P}_k(\mathcal{H})$ onto itself maps bijectively $\mathcal{P}_{qk}(\mathcal{H})$ onto itself for $q \in \mathbb{N}$.
  \end{cor}

  \begin{remark}
  Note, that Wigner's theorem immediately follows from preserving the orthogonality relation and the properties of a spectrum preserver. Indeed, preserving the orthogonality relation of rank-1 projectors implies that a Schatten decomposition $\sum_i \lambda_i P_i$ of a hermitian operator is mapped to another Schatten decomposition with the same spectrum. Such a map is therefore a spectrum preserver and hence has a form $\Phi(X) = M X M^{-1}$ or $\Phi(X) = M X^{\rm t} M^{-1}$. Finally,  preservation of orthogonality  implies that $M$ is unitary.
  \end{remark}

\begin{prop} \label{rec}
Assume, that $k = n \mod l$. If any linear map $\Psi: \BHH(\mathcal{H}) \to \BHH(\mathcal{H})$ transforming $\mathcal{P}_k(\mathcal{H})$ onto itself bijectively is of the form (\ref{!}), then also each map $\Phi: \BHH(\mathcal{H}) \to \BHH(\mathcal{H})$ transforming $\mathcal{P}_l(\mathcal{H})$ onto itself bijectively is of the form (\ref{!}).

   \proof Let $\Phi: \BHH(\mathcal{H}) \to \BHH(\mathcal{H})$ maps bijectively $\mathcal{P}_l(\mathcal{H})$ onto itself. Let $n = k + q \cdot l$. Then due to Corollary \ref{qk}, for any $P \in \mathcal{P}_{q l}(\mathcal{H})$ one has  $\Phi(P) \in \mathcal{P}_{q l}(\mathcal{H})$. Now, any projector from $\mathcal{P}_{q l}(\mathcal{H})$ may be written as $I - P_k$ with $P_k \in \mathcal{P}_k(\mathcal{H})$ and hence
\begin{equation}\label{a}
  \Phi(I - P_k) = \Phi(I) - \Phi(P_k) =: I - Q_k \ ,
\end{equation}
for some $Q_k \in \mathcal{P}_{k}(\mathcal{H})$. One has therefore
\begin{equation}
  Q_k = \Phi(P_k) - \Phi(I) + I \ .
\end{equation}
The above relation defines a map $P_k \rightarrow Q_k$ transforming $\mathcal{P}_k(\mathcal{H})$ onto itself bijectively, so due to our assumption it can be written as
\begin{equation}
  Q_k = U \widetilde{P}_k U^\dagger\ .
\end{equation}
where $\widetilde{X}=X$ or $\widetilde{X}=X^{\rm t}$. Finally
\begin{equation}
 \Phi(P_k) = \Phi(I) - I + U \widetilde{P}_k U^\dagger\ ,
\end{equation}
and by linearity it may be extended to the following linear map on $\BH$
\begin{equation}\label{}
  \Phi(X) = \frac 1k[\Phi(I) - I]{\rm Tr X} + U\widetilde{X}U^\dagger\ .
\end{equation}
In particular if $X = P_l \in \mathcal{P}_{l}(\mathcal{H})$ one has
\begin{equation}\label{}
  \Phi(P_l) = \frac lk [\Phi(I) - I] + Q_l =: D + Q_l\ ,
\end{equation}
where $Q_l = U\widetilde{P}_lU^\dagger\in \mathcal{P}_{l}(\mathcal{H})$. To complete the proof we need to show that $D=0$. Since $D+Q_l$ is a projector one has $(D + Q_l)^2 = D+ Q_l $ and hence
\begin{equation}\label{}
  DQ_l + Q_l D = D - D^2 .
\end{equation}
Taking into account that $\Phi$ is trace-preserving one has ${\rm Tr}D = 0$ which implies $2{\rm Tr}(D Q_l) = - {\rm Tr}D^2$ for all $Q_l \in  \mathcal{P}_{l}(\mathcal{H})$. Now, if $Q_l$ is a projector on the $l$-dim. subspace spanned by eigenvectors of $D$ corresponding to $l$ largest eigenvalues of $D$, i.e $d_1 \geq d_2 \geq \ldots \geq d_l$,  then $2{\rm Tr} (Q D) = 2\sum_{i=1}^l d_i = - {\rm Tr}D^2 \leq 0$ but since $D$ is traceless one has $\sum_{i=1}^l d_i \geq 0$ which proves that $D=0$. \hfill $\Box$

\end{prop}

The main result of the paper is provided by the following

\begin{thm} If $n$ is prime, then any $\Phi$  mapping surjectively rank-$k$ projectors into  rank-$k$ projectors is of the form (\ref{!}).

\proof
Let us define a sequence via formula $k_{i+1} = n \mod k_i$ and $ k_0 = k$. This is strictly decreasing, finite sequence, and it terminates at $0$ and let $0=n \mod k_*$, that is, the sequence reads $\{k_0=k> k_1 > k_2 > \ldots > k_* > 0\}$.  Due to the Proposition \ref{rec}, if any $\Psi: \BHH(\mathcal{H}) \to \BHH(\mathcal{H})$ maps bijectively $\mathcal{P}_{k_{i+1}}(\mathcal{H})$ onto itself is of the form (\ref{!}), then any $\Psi: \BHH(\mathcal{H}) \to \BHH(\mathcal{H})$ mapping bijectively $\mathcal{P}_{k_i}(\mathcal{H})$ onto itself is of the form (\ref{!}) as well.  Now, due to the Wigner's theorem, if $k_*=1$ then for any $k_i> k_*$ in the sequence, any $\Psi: \BHH(\mathcal{H}) \to \BHH(\mathcal{H})$ mapping bijectively $\mathcal{P}_{k_i}(\mathcal{H})$ onto itself is of the form (\ref{!}), in particular $k_0=k$. Note, that $k_*$ is by construction a divisor of $n$ and hence if $n$ is prime, then $k_*=1$ which complete the proof. \hfill $\Box$
\end{thm}

\vspace{.2cm}

Observe, that $n=q_i k_i + k_{i+1}$ and if for some number $d$ one has $d|n$ and $d|k_i$ then $d|k_{i+1}$, so the common divisors of $n$ and $k_0=k$  are also the common divisors of all elements of the sequence. Thus this method does not give a conclusive answer if the starting point $k_0=k$ is not relatively prime to $n$. If $n$ and $k$ are relatively prime then a conclusive answer is not guaranteed. Indeed, if $k=3$ and $n=10$ one has the conclusive answer, but for $k=3$ and $n=8$ the metod does not give a conclusive answer. Having in mind, that $k_*$ is by construction a divisor of $n$, one has the following 

\begin{remark} \label{k|n} To complete the characterisation of surjective maps from $\mathcal{P}_{k}(\mathcal{H})$ into $\mathcal{P}_{k}(\mathcal{H})$ it is enough to characterise these maps for $k|n$.
\end{remark}


%
%
%

\begin{remark} Let $n=2k$. Consider
\begin{equation}\label{!!}
  \Phi(X) = \frac{\mathbb{I}_{2k}}{k} {\rm Tr} X - X \ .
\end{equation}
It is clear that $\Phi$ maps rank-$k$ projectors into rank-$k$ projectors but evidently it does not have the form (\ref{!}). Indeed, if $P$ is rank-1 projector then $\Phi(P)$ has rank $n-1$ and hence it is not rank-1 projector. Note, that $\Phi^{-1}=\Phi$.
\end{remark}

Such maps are the already defined (\ref{R}) and because $\mathcal{P}_k(\mathcal{H})$ and $\mathcal{P}_{n-k}(\mathcal{H})$ are the same set, these maps are  endomorphisms. This encourages us to claim that these are the only additional endomorphisms. Therefore we pose the following

\begin{con} \label{n=2k}
  Let $n=2k$ and $\Phi: \mathcal{P}_{k}(\mathcal{H}) \ra \mathcal{P}_{k}(\mathcal{H})$ surjectively. Then the map is of the form $\Phi$ or $\Phi \circ R_k$, where $\Phi$ is of the form (\ref{!}) and $R_k$ is defined by (\ref{R}).
\end{con}

This conjecture allows to perform  perfect characterization of surjective maps $\mathcal{P}_{k}(\mathcal{H}) \ra \mathcal{P}_{k}(\mathcal{H})$

\begin{prop} Let us assume that Conjecture 1 is true. If  $k|n$, $n/k>2$ and $\Phi: \mathcal{P}_{k}(\mathcal{H}) \ra \mathcal{P}_{k}(\mathcal{H})$ is surjective, then $\Phi$ has the form (\ref{!}).

 \proof Let $\{e_i\}_{i=0}^{n-1}$ be the standard basis of $\mathcal{H}$ and let $P_i$ be a rank-$k$ projector defined by
 
$$ P_i = \sum_{j=0}^{k-1} |e_{ik+j} \rangle \langle e_{ik+j}|\ , \ \ i=0,1,\ldots,\frac nk-1 \ .$$ 
It is clear that $P_i$ and $P_j$ are mutually orthogonal for $i\neq j$. Now, due to  the Proposition \ref{ort}, a set of projectors $\{P_i\}$ is mapped to the set of pairwise orthogonal rank-$k$ projectors. Now, since 
$$ P_0 + \ldots + P_{n/k} =  \mathbb{I} =  \Phi(P_0) + \ldots + \Phi(P_{n/k})   $$
projectors $\{ \Phi(P_0) , \ldots , \Phi(P_{n/k}) \}$ are  unitarly equivalent to $ \{ P_0 , \ldots , P_{n/k} \}$, that is, $\Phi(P_i) = UP_i U^\dagger$ for some unitary $U$.  Without loosing generality we may assume that $U=\mathbb{I}$, that is, $\Phi$ maps $P_i$ to $P_i$. Let $V_i = \mathrm{span} \{ e_{k\cdot i}, \dots e_{k\cdot i +(k-1)}\}$ be the range of $P_i$ and let $\mathbb{I}_{ij}$ be the projector onto the subspace $V_i \oplus V_j$. Let

$$\Phi_{ij} := \mathbb{I}_{ij}\, \Phi\, \mathbb{I}_{ij} , $$
be a restriction of $\Phi$ to the subspace $V_i \oplus V_j$. By the Conjecture \ref{n=2k} it is of the form
$\Phi_{ij}(X)=U_{ij}\widetilde{\Phi}_{ij}(X)U_{ij}^\dagger$, where  $\widetilde{\Phi}_{ij}$ maps $X$ to  $X, X^t, \frac 1k \mathbb{I}_{ij} \Tr X-X$ or $\frac 1k \mathbb{I}_{ij} \Tr X-X^t$.
One has that for all $i,j$  $\Phi_{ij} (P_i) = P_i$ and $\Phi_{ij} (P_j) = P_j$, hence for all $\alpha,\beta$
\begin{displaymath}
 \left[ \begin{array}{cc} \alpha P_i & \\ & \beta P_j \end{array} \right] = 
 U_{ij} \left[ \begin{array}{cc} \alpha P_i & \\ & \beta P_j \end{array} \right] U_{ij}^\dagger = 
 \left[ \begin{array}{cc} A & B \\ C & D \end{array} \right]
 \left[ \begin{array}{cc} \alpha P_i & \\ & \beta P_j \end{array} \right]
 \left[ \begin{array}{cc} A & B \\ C & D \end{array} \right]^\dagger
\end{displaymath}
what implies that $U_{ij}$ is block-diagonal if $\widetilde{\Phi}_{ij}(X) =  X$ or $X^t$ and $U_{ij}$ is block-antidiagonal if $\widetilde{\Phi}_{ij}(X) =  \frac 1k \mathbb{I}_{ij} \Tr X-X$ or $\frac 1k \mathbb{I}_{ij} \Tr X-X^t$.

 Moreover, if $\psi \in V_i$ then  
 
$$ \Phi_{ij}(|\psi\rangle\langle\psi|) = \mathbb{I}_{ij} \Phi(|\psi\rangle\langle\psi|) $$ 
does not depend on $j$, and similarly if  $\phi \in V_j$ then

$$ \Phi_{ij}(|\phi\rangle\langle\phi|) = \mathbb{I}_{ij} \Phi(|\phi\rangle\langle\phi|) $$ 
does not depend on $i$.  It follows, that all $\Phi_{ij}$ does nor depend on $i,j$ and hence $\widetilde{\Phi}_{ij}(X)=X$ or $X^t$ (otherwise it could not  give the same result for different $j$s if $n/k>2$, as the reduction map has the information about the trace of the second block). Now we know that all $U_{ij}$s are block-diagonal and again, because $\Phi_{ij}(|\Psi\rangle\langle\Psi|)$ has to give the same result for all $j$s one gets that $U_{ij} = U_i \oplus U_j$.

 Finally we get that 
 
$$ \Phi(X) =  U \, X \, U^\dagger \ \ \ \mbox{or} \ \ \ 
 \Phi(X) =   U\,  X^t\,  U^\dagger  $$
with $U =  \bigoplus_{i=1}^{n/k} U_i$, which ends the proof. \hfill $\square$

\end{prop}

\section{Conclusions}

Let us summarise the paper by the following remarks:

\begin{remark} Let us observe that if we relax the condition that the map $\Phi$ is invertible then one may have maps from $\mathcal{P}_k(\Cn)$ to $\mathcal{P}_l(\Cn)$  with $l \neq n-k$. The well known example is provided by the Breuer-Hall map $\Phi_{\rm BH} : M_{2n}(\mathbb{C}) \ra  M_{2n}(\mathbb{C})$ defined as follows \cite{Breuer,Breuer-C,Hall}
\begin{equation}\label{}
  \Phi_{\rm BH}(X) = \frac{1}{2(n-1)} \left( \mathbb{I}_{2n} {\rm Tr} X - X - U X^{\rm t} U^\dagger \right) ,
\end{equation}
where $U$ is an arbitrary anti-symmetric $2n \times 2n$ matrix. $\Phi_{\rm BH}$ maps rank-1 projectors into projectors of rank $2(n-1)$. It is evident that $\Phi_{\rm BH}$ is not invertible.

\end{remark}

\begin{remark} Let us observe that maps (\ref{!}) are characterized by the following property: $\Phi$ is positive and trace-preserving and $\Phi^{-1} = \Phi^*$, where the dual map $\Phi^*$ is defined by $(X,\Phi(Y))_{\rm HS} = (\Phi^*(X),Y)_{\rm HS}$. Interestingly, the map (\ref{!!}) is characterized by the following property:  $\Phi$ is trace-preserving and $\Phi^{-1} = \Phi^*=\Phi$. It means that both (\ref{!}) and (\ref{!!}) are isometries with respect Hilbert-Schmidt product and hence the corresponding eigenvalues satisfy $|\lambda_i|=1$.

On the other hand, any rank-1 projector $P$ can be decomposed as a combination o rank-$k$ projectors commuting with $P$ and hence any hermitian operator can be decomposed as a combination of $n$ commuting rank-$k$ projectors. This combination is mapped into another combination of rank-$k$ projectors (in general not commuting) with the same coefficients and hence there exists a common upper bound for the maximal eigenvalue of all invertible maps mapping $\mathcal{P}_k(\mathcal{H})$ to itself. While such maps form a group, their eigenvalues has to lie therefore on the unit circle and at least one of them is equal $1$. We stress that this property is not equivalent to being a Hilbert-Schmidt isometry. If one could prove that for any invertible map $\Phi$ mapping $\mathcal{P}_k(\mathcal{H})$ to itself its adjoint map $\Phi^*$ has the same property, then it would imply that this class is a subclass of Hilbert-Schmidt isometries.
\end{remark}

\begin{remark}  The crucial difference between (\ref{!}) and (\ref{!!}) is positivity. Let us recall \cite{Bhatia} that a trace-preserving map is positive iff
\begin{equation}\label{11}
  || \Phi(X)||_1 \leq ||X||_1 \ ,
\end{equation}
for all self-adjoint elements $X$. It is clear that (\ref{!!}) is not positive and hence violates (\ref{11}). Interestingly, for $X$  traceless one has
\begin{equation}\label{}
  ||\Phi(X)||_1 = ||-X||_1 = ||X||_1 \ .
\end{equation}
In particular taking two density operators $\rho_1$ and $\rho_2$ one has
\begin{equation}\label{}
  ||\Phi(\rho_1-\rho_2)||_1 = ||\rho_1 - \rho_2||_1 \ ,
\end{equation}
which means that (\ref{!}) preserves the distinguishability between arbitrary quantum states. Clearly, (\ref{!}) enjoys the same property.
\end{remark}

Finally, it is hoped that presented result finds applications in quantum information theory (e.g. entanglement detection) and the analysis of symmetries of quantum systems (e.g the evolution of quantum systems).

\section*{Acknowledgements}

We thank anonymous referees for valuable comments.



\begin{thebibliography}{1} \bibliographystyle{plain}


\bibitem{Wigner} E. P. Wigner, Group Theory, Academic Press, New York, 1959.

\bibitem{Lahti} G. Cassinelli, E. De Vito, P. J. Lahti, A. Levrero, {\em The Theory of Symmetry Actions in Quantum Mechanics with an Application to the Galilei Group}, Springer 2004.

\bibitem{Simon} R. Simon, N. Mukunda, S. Chaturvedi, and V. Srinivasan, Two elementary proofs of the Wigner theorem on symmetry in quantum mechanics, Phys. Lett. A {\bf 372}, 17 (2008): {\em ibidem} {\bf 378},13 (2014).

\bibitem{Paulsen} V. Paulsen, {\it Completely Bounded Maps and Operator Algebras}, Cambridge University Press, 2003.

\bibitem{Bhatia} R. Bhatia, {\em Positive Definite Matrices}, (Princeton University Press, 2006).

\bibitem{Werner} F. Buscemi, G. M. D'Ariano, M. Keyl, P. Perinotti, R. Werner,  Clean Positive Operator Valued Measures, 	J. Math. Phys. {\bf 46}, 082109 (2005)


\bibitem{Adam} M. Mozrzymas, A. Rutkowski, and M. Studzi\'nski, Using non-positive maps to characterize entanglement witnesses, J. Phys. A: Math. Theor. {\bf 48}, 395302 (2015).

\bibitem{LIN} Chi-Kwong Li  and Nam-Kiu Tsing, Linear Preserver Problems:
A Brief introduction and Some Special Techniques, Lin. Alg. Appl. {\bf 162-164}, 217 (1992).

\bibitem{Frobenius} G. Frobenius, \"Uber die Darstellung der endlichen Gruppen durch lineare
Substitutionen, Sitzungsber. Deutsch. Akad. Wiss. Berlin, 1897, pp. 994-1015.

\bibitem{NIL} P. Botta, S. Pierce, and W. Watkins, Linear transformations that preserve the nilpotent matrices, Pacific J. Math. {\bf 104}, 39 (1983).

\bibitem{MM} M. Marciniak {\em On extremal positive maps acting between type I factors} Noncommutative Harmonic Analysis with Applications to Probability II, Banach Center Publications, vol {\bf 69} Institute of Mathematics, Polish Academy of Science, Warsaw 2010

\bibitem{Breuer} H.-P. Breuer, {\em Optimal Entanglement Criterion for Mixed Quantum States}, Phys. Rev. Lett. {\bf 97}, 0805001
(2006).

\bibitem{Breuer-C} H.-P. Breuer, {\em Separability criteria and bounds for entanglement measures}, J. Phys. A: Math. Gen. {\bf 39}, 11847 (2006).


\bibitem{Hall} W. Hall, {\em A new criterion for indecomposability of positive maps},  J. Phys. A: Math. Gen. {\bf 39}, (2006)
14119.

\end{thebibliography}
\end{document}